\documentclass[10pt,a4paper]{article}
\usepackage{opex3}
\usepackage{graphicx}
\usepackage{hyperref}
\usepackage{amssymb}
\newcommand{\doi}[2]{\href{http://dx.doi.org/#1}{#2}}
\newcommand{\PRA}[3]{\href{http://dx.doi.org/10.1103/PhysRevA.#1.#2}{Phys.\ Rev.\ A \textbf{#1}, #2 (#3)}}
\newcommand{\PRAR}[3]{\href{http://dx.doi.org/10.1103/PhysRevA.#1.#2}{Phys.\ Rev.\ A \textbf{#1}, #2(R) (#3)}}
\newcommand{\PRL}[3]{\href{http://dx.doi.org/10.1103/PhysRevLett.#1.#2}{Phys.\ Rev.\ Lett.\ \textbf{#1}, #2 (#3)}}
\input epsf

\begin{document}

\title{Dark dynamic acousto-optic ring lattices for ultracold atoms}

\author{N.\ Houston, E.\ Riis, A.\ S.\ Arnold}
\address{SUPA, Dept.\ of Physics, University of Strathclyde, Glasgow G4 0NG, UK}

\begin{abstract}
We demonstrate the optical generation of dynamic dark optical ring lattices, which do not require Laguerre-Gauss beams, large optical coherence lengths
or interferometric stability. Simple control signals lead to spatial modulation and reproducible rotation, offering manifold possibilities for complex
dynamic ring lattices. In conjunction with a magnetic trap, these scanned 2D intensity distributions from a single laser beam will enable precision
trapping and manipulation of ultracold species using blue-detuned light. The technique is ideal for azimuthal ratchet, Mott insulator and persistent
current experiments with quantum degenerate gases.
\end{abstract}

\ocis{(020.7010) Trapping; (230.1040) Acousto-optical devices}%; (230.6120) Spatial light modulators}

%Clarify last section and more physics in first section?
\section{Introduction}
Laser cooling and the subsequent attainment of %bosonic and fermionic
quantum degenerate gases has enabled exquisite control over atoms. Coherent atom-optical manipulation of coherent atomic gases, typically with magnetic
fields and/or far-detuned lasers, is now a worldwide phenomenon \cite{general}. There are a plethora of available atom trapping geometries and rings are
of particular interest because they enable precision Sagnac interferometry and detailed studies of superfluidity, plus the periodic boundary conditions
afford simple modeling of the system.

Bose-condensed atoms have recently been obtained in magnetic ring geometries with $1-50\,$mm diameters \cite{gup} and persistent currents \cite{phill}
have been observed in a small-scale condensate ring trap \cite{raman}. Ultracold atom ring traps based on electrostatic potentials \cite{mabuchi} and
induced currents \cite{griff} have also been proposed, and RF dressed rings have been experimentally realised \cite{dressexp}.

Confinement within optical lattice geometries offers further possibilities; potentially as a tool for performing universal simulation of quantum dynamics
\cite{Cirac}, and for realisation of Feynman's ideas of quantum logic \cite{Feynman}, thus forming a promising basis for quantum computation
\cite{Adams}. Optical lattices enable condensed matter physics investigations, such as the quantum phase transition from superfluidity to Mott insulator
\cite{Mott} and the realisation of Josephson junction arrays \cite{Josephson}. The combination of rings and lattices to form ring lattices with
rotational symmetry and periodic boundary conditions is particular beneficial for e.g.\ studies of solitons, quantum many particle systems, entanglement,
Mott transitions and persistent currents \cite{ringlatth}.

Ring lattice potentials utilise far-detuned optical dipole beams, which can be clearly divided into `bright' or `dark' lattices: atoms are attracted to
dark (bright) spatial regions of the time-averaged optical potential if one uses light blue-detuned (red-detuned) from an atomic resonance. Blue-detuned
light is preferable for optical manipulation, as atoms trapped in low intensity light experience lower photon scattering (i.e.\ heating) rates, energy
level shifts, and light assisted collisional losses \cite{Davidson}. Blue-detuned (`dark') trapping is likely to be necessary for lattice based quantum
computation, to ensure robustness against decoherence \cite{Adams}.

Dark optical ring lattices \cite{Hill} have been previously realised using copropagating superpositions of Laguerre-Gauss (LG) laser beams \cite{Ferris}.
Complex optics (spatial light modulators, SLMs) were used to generate separate static LG beams, which were then frequency shifted by independent
acousto-optic modulators (AOMs). For a dynamic lattice the LG beams must be recombined on a beamsplitter, leading to power loss and a requirement for
interferometric relative beam stability to ensure reproducible long term experiments. An alternative way to produce smooth time-dependent lattices is to
update dynamically the SLM pattern; however update rates tend to be slow, and for the faster binary SLMs additional control algorithms must also be used,
due to SLM interframe artifacts \cite{Foot}. %In combination with unwanted optical interference between sites, these factors can detrimentally affect ring
%lattices, particularly dark lattices generated from light detuned towards the blue of an atomic resonance.

Bright ring lattices can be directly generated with AOMs. The acoustic mode frequency and amplitude in an AOM determine the beam deflection angle and
intensity respectively, allowing spatial control \cite{KetterleSE}. By scanning the position of a laser beam in one or two dimensions it is possible to
create a BEC mirror \cite{Ertmer}, condensate surface excitations and multiple-site traps \cite{KetterleSE}, or `stir' a condensate to produce vortices
\cite{dalibard}. Two-dimensional \textit{bright} ring lattice potentials using scanned red-detuned laser beams have been suggested \cite{Bouyer},
optically realised \cite{Matt} and experimentally implemented for storing and splitting Bose-Einstein condensates \cite{Boshier}.

Here we utilize two-dimensional AOM beam scanning \cite{KetterleSE,Matt,Boshier} to optically demonstrate simple \textit{dark} optical ring
lattice potentials for use with ultracold atoms or BECs, without the need for relative laser beam phase coherence required in reference
\cite{Ferris}. In BEC experiments using destructive imaging, the requirement of a ring lattice with a well defined shot-to-shot angular phase is
crucial and our potentials can be reproducibly rotated around the beam axis and also spatially modulated. Moreover, if additional confinement is
provided by a magnetic field (section~\ref{atopt}), a blue-detuned realization of the dark ring lattice will be essentially decoherence free
\cite{Adams} and highly adaptable -- ideal for Mott insulator \cite{Mott}, persistent current \cite{phill} and azimuthal ratchet \cite{renzoni}
experiments involving quantum degenerate gases. We will discuss the ring lattice theory, describe the experiment, then place our results in an
atom optics context.

\section{Theory}
\begin{figure}[!b]
\begin{center}\includegraphics[width=.8\columnwidth]{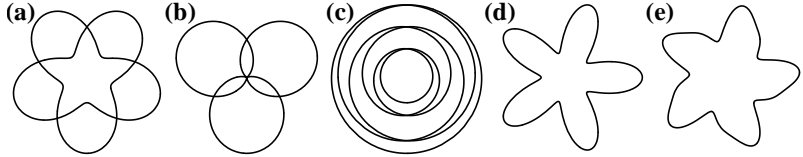}\end{center}\vspace{0mm}
\vspace{-4mm}\caption{\label{Fig:1}Parametric curves described by Eq. (1) with $R=A+B\sin\omega_2t$ and $\{A/B,\,\omega_2/\omega_1\}$ set to:
$\{1/2,\,5/2\}$ (a); $\{1,\,3/2\}$ (b); $\{1/2,\,1/6\}$ (c); $\{1/2,\,5\}$ (d). In (e) the angular symmetry of the lattice is broken using
$R=A(1+\sin(\omega_2 t)/5+\sin(2\omega_2 t)/20).$}
\end{figure}

%Parametric equations are often used in mathematics to describe families of curves with rotational symmetry, such as hypotrochoids and epitrochoids
%(commonly recognised as `Spirograph' curves), or Lissajous figures.
We consider ring-lattice potentials based on a laser beam following a 2D path of the form:
 \begin{equation}\label{veceq}
    \{x, y\}= R \{\sin(\omega_1 t),\cos(\omega_1 t)\}\\
 \end{equation}
where $R(\omega_2 t+\phi)$ is an arbitrary function with period $2\pi/\omega_2$ (i.e. $R$ is comprised of Fourier components at angular frequencies that
are integer multiples of $\omega_2)$. The overall period of the parametric function is $T=2\pi/$gcd$(\omega_1,\omega_2)$ (gcd=greatest common divisor).
Equation~(\ref{veceq}) forms a large family of rotationally symmetric `flower like' curves (Figure~\ref{Fig:1}) defined by the Fourier components of $R.$
The frequency ratio $\omega_2/\omega_1$ is related to the number of `petals' in the curve and any curve can be directly rotationally modulated using the
overall phase of $R$, i.e.\ $\phi.$

Half integer values of $\omega_2/\omega_1$ produce curves with a ring lattice profile (Figure \ref{Fig:1}(a)), integer values of $\omega_2/\omega_1$
produce curves which, when combined with a magnetic quadrupole potential (section~\ref{atopt}) create a hybrid magneto-optic ring lattice (Figure
\ref{Fig:1}(d)) and it is also possible to make patterns with broken angular symmetry for ring ratchet experiments (Figure \ref{Fig:1}(e)). Additionally,
by adjusting the phase $\phi$ of the amplitude modulation $R$ during the circular beam modulation (at $\omega_1$), angular rotation or angular modulation
of the curve can be achieved.
%The parametric equation (\ref{veceq}) is therefore an ideal and highly adaptable model with which to control and rotate optical ring lattices.

Laser beam deflection from an AOM is proportional to the RF drive frequency.  The deflection can be modulated by altering the drive RF frequency, which
can be achieved by varying the voltage driving a voltage controlled oscillator (VCO). Thus AOM deflection is synchronised to the VCO input voltage. If a
laser beam is passed through two perpendicular AOMs, with corresponding VCOs driven by the parametric signal Eq.\ (\ref{veceq}), the variation in the
vertical and horizontal deflection angles will cause the beam to trace out the corresponding parametric curves. If this variation is rapid compared to
typical atomic velocities, ultracold species will effectively experience only the time-averaged geometry of the beam trace \cite{top}, allowing both
trapping and spatial control of atoms or BECs within the scanned beam.

%We demonstrate that by driving two crossed AOMs with synchronised signals of the form , these parametric curves can be realised as dynamic two
%dimensional intensity patterns, such as optical ring lattices, from a single laser beam.

\section{Experiment}

The design used to generate the required VCO control signals is shown schematically in Figure 2. The circuit is able to operate from a single `master'
synthesized signal generator (SSG$_{\rm M}$, e.g.\ Agilent 33220A or SRS DS345) by taking advantage of the 10MHz clock output, to synchronise a second
`slave' SSG$_{\rm S}$ (or control circuitry \cite{circuit}). SSG$_{\rm M}$ is used for radial modulation of the circular beam path provided by SSG$_{\rm
S}$. We have used analogue multipliers because the amplitude modulation function of a typical SSG has limited bandwidth.

\begin{figure}[!h]
\begin{center}\includegraphics[width=\columnwidth]{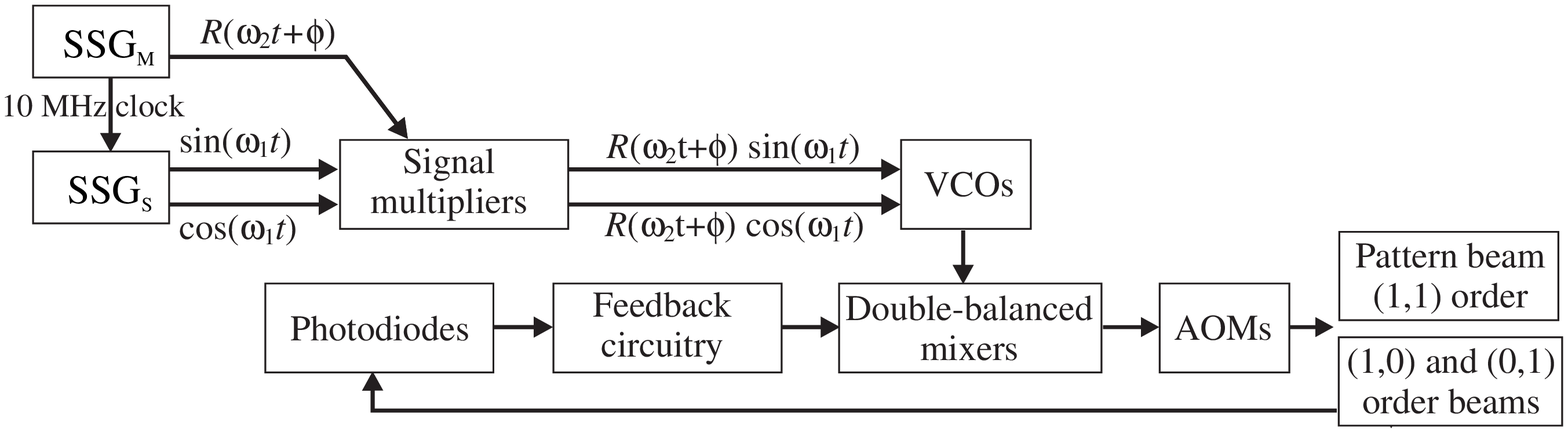}\end{center}\vspace{0mm}
\vspace{-2mm}\caption{\label{Fig:2} The signal generation circuit is based on two synthesized signal generators (SSG). Our slave synthesizer SSG$_{\rm
S}$ is actually a custom-made circuit based on low-cost ICs \cite{circuit}.}
\end{figure}

The multiplier outputs (Eq.\ ~\ref{veceq}) are then fed to two independent VCOs to provide the RF signals for driving two $110\,$MHz AOMs.
Voltage controlled attenuators (VCAs) or double-balanced mixers are then used for control over RF power, and thus beam intensity, during the
beam scan. Our base frequency is currently set to $\omega_1=2\pi\times 10\,$kHz (a subharmonic of the SSG$_{\rm M}$ clock) for convenience,
however with sufficient VCO bandwidth, this could easily be increased by more than a factor of ten.
%Thus, and at present w generate patterns with integer (half integer) values of $\omega_2/\omega_1$ with a period of $100\,\mu$s ($200\,\mu$s).

\begin{figure}[!h]
\begin{center}\includegraphics[width=.86\columnwidth]{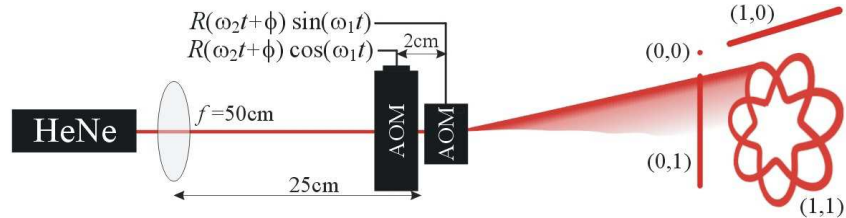}\end{center}\vspace{-3mm}
\caption{\label{Fig:3} Experimental setup for generation of optical ring lattices. Synchronised variation in AOM deflection angles causes the beam to
trace out the ring lattice shown.}
\end{figure}

Figure~\ref{Fig:3} shows the experimental setup. A helium-neon laser beam is focused through two perpendicular AOMs, which are placed close to
each other to minimise output distortion and scan asymmetry. Each AOM produces a spread of diffractive orders which results in a grid of
diffracted beams at the output.
%, so an aperture (not shown) is used to select the beam corresponding to each AOM's +1 diffraction order.

\begin{figure}[!h]
\begin{center}
 \includegraphics[width=.7\columnwidth]{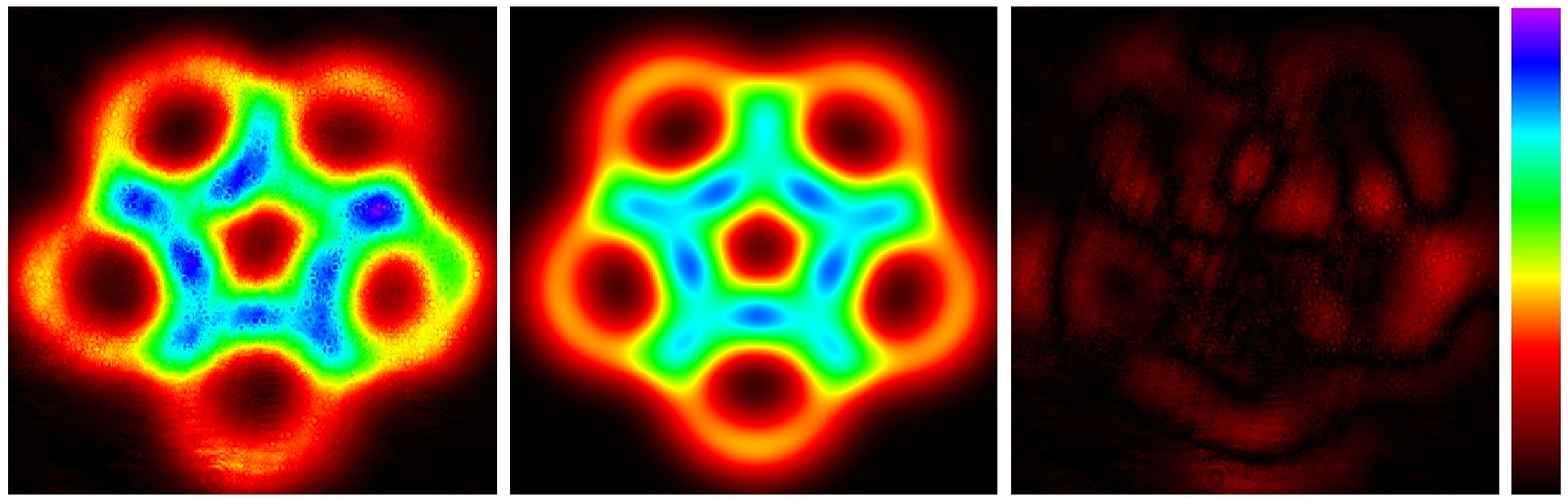}\vspace{1mm}
 \includegraphics[width=.7\columnwidth]{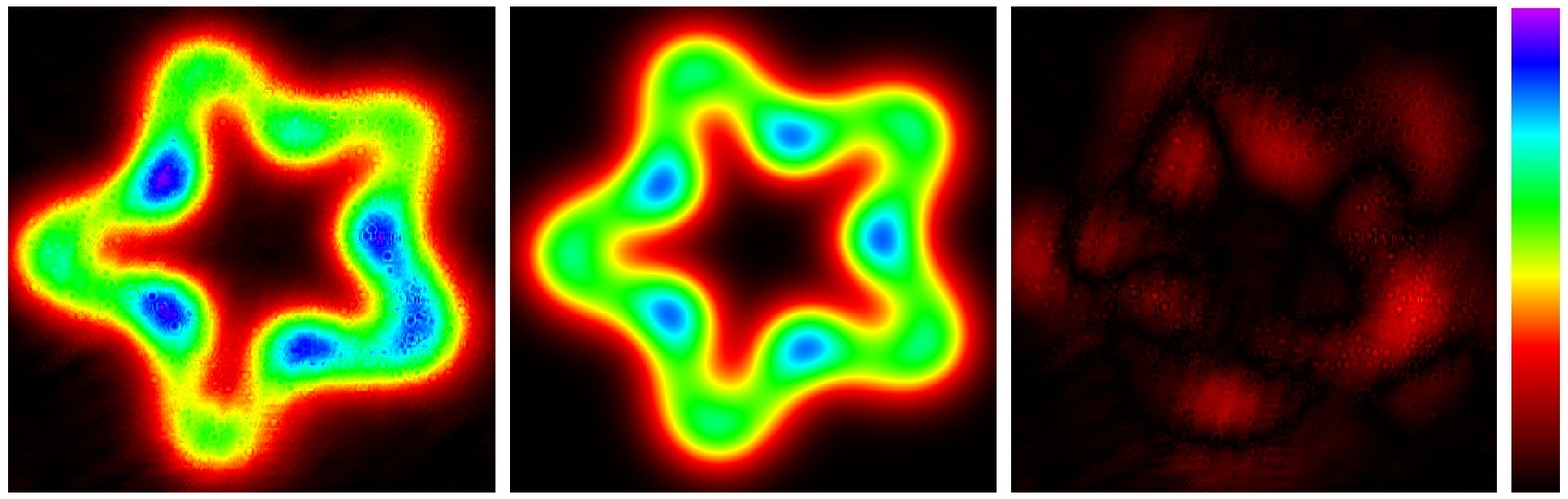}
\end{center} \vspace{0mm} \caption{\label{Fig:4} Experimental relative intensity distributions (area $\approx(4\,$mm$)^2$, exposure $1\,$ms),
corresponding least-squares theoretical fits using Eq.~\ref{rast} and fit residues. Click
\href{http://www.photonics.phys.strath.ac.uk/AtomOptics/AOFerris.html}{here} for optical lattice movies comparing experimental and theoretical
rotation and amplitude modulation. The static laser beam has a waist ($e^{-2}$ radius) $\approx300\,\mu$m.}
\end{figure}

%\section{Ring lattice results}
Figure \ref{Fig:4} shows experimental results for 5-site optical ring lattices realised using this technique, as well as comparison to 10-parameter 2D
least-squares fits from a simple theoretical model based on a Gaussian beam with $x$ and $y$ beam waists of $w_x$ and $w_y$ respectively, scanned in the
$xy$ plane yielding a time-averaged intensity:
\begin{equation}
\overline{I(x,y)}= \frac{I_0\,s \omega_1}{2 \pi} \int_{0}^{\frac{2 \pi}{s\omega_1}} {\exp\left[-2 (x-x_{r}(t))^2/{w_x}^{2}-2
(y-y_r(t))^2/{w_y}^{2}\right] dt;} \label{rast}
\end{equation}
where $\{x_r,y_r\}=\{x_0-(r_x+A_x \sin(5 s \omega_1 t+\phi)) \sin(\omega_1 t),y_0-(r_y+A_y \sin(5 s \omega_1 t+\phi)) \cos(\omega_1 t)\}$ is the position
of the center of the scanned beam at time $t.$ The parameter $s$ is set to $\frac{1}{2}$ or 1 for a closed optical ring lattice (cf. Fig.~\ref{Fig:1}
(a)) or open optical ring lattice (cf. Fig.~\ref{Fig:1} (d)) respectively.

As the contribution to the time-averaged intensity is inversely proportional to the velocity of the center of the scanned beam, there is a
radial intensity gradient due to the deflected beam spending longer in the center of the pattern. This gradient does not affect the quality of
the trapping potential for e.g. Mott insulator experiments, as long as the pattern has a high level of rotational symmetry. It is therefore
important to scan a dipole beam with an aspect ratio as close as possible to 1, with a stable beam intensity during the scan process. To this
end we have implemented a feedback mechanism (`noise-eater') to the VCA control signals based on error signals generated from the unused (1,0)
and (0,1) AOM deflection orders (Fig.~\ref{Fig:3}). The advantage of feedback, as opposed to recorded feedforward \cite{Matt}, is that it
immediately adapts to beam intensity noise due to environmental changes or changes to the amplitude modulation function. Although high bandwidth
is required for feedback, we have already reduced intensity amplitude noise to $5\%$ rms, and anticipate future improvements.

\section{Atom optics}\label{atopt}
Whilst there has been growing theoretical interest in ring lattices \cite{ringlatth}, the closest experiment to a dark BEC ring lattice has consisted of
three sites produced by variable intensity light traversing a static mechanical aperture \cite{anderson}. The azimuthal and central barriers of this
3-site ring lattice must be altered together. In contrast the azimuthal and central barrier of our $N$-site optical potential can be altered
independently. Additionally, as the azimuthal lattice angle is purely controlled by the phase $\phi$ of the VCO amplitude modulation, the ring lattice
can be reproducibly rotated, even for BECs with experimental production times of minutes.

\begin{figure}[!h]
\begin{center}
\includegraphics[width=\columnwidth]{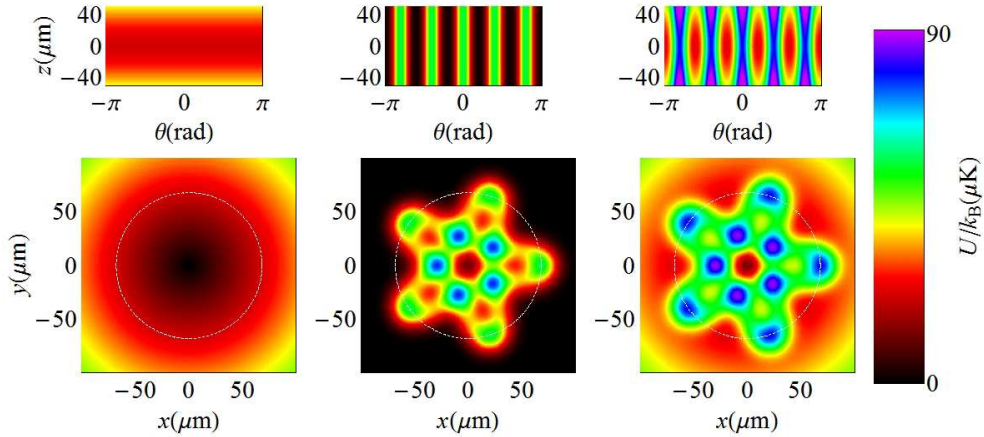}
 \end{center}\vspace{-3mm}
\caption{\label{Fig:5} Magnetic (left), optical (center) and hybrid magneto-optical potentials (right). The upper images show axial-azimuthal
slices indicated by the dashed white circles.}
\end{figure}

We now give a brief illustration of experimentally realistic parameters for a 5-site optical ring lattice. The scanned laser beam leads to an
optical dipole trap with atomic scattering rate (Hz) and depth (J) approximately given by $S\approx\Gamma \overline{I}/(8 I_{\rm S}
{\Delta_\Gamma}^2)$ and $U\approx \hbar \Gamma \overline{I}/(8{\Delta_\Gamma} I_{\rm S})$ respectively, where $\overline{I}$ and
$\Delta_\Gamma=(\omega-\omega_0)/\Gamma$ are the time-averaged spatial intensity and laser detuning from the atomic transition (in linewidths).
A very useful, species independent parameter for blue-detuned dipole traps is $S/T=k_{\rm B}/(\hbar \Delta_\Gamma)$ - the \textit{maximum}
scattering rate experienced by an atom with total energy $U=k_{\rm B} T$ as a function of the dipole trap depth. For the $780\,$nm D2 transition
in $^{87}$Rb, $\Gamma=2\pi\times 6\,$MHz and $I_{\rm S}=16.7\,$W/m$^2$. If we use a dipole trap laser with wavelength $\lambda=765\,$nm
$(\Delta_\Gamma\approx 1.3\times10^6)$ we have $S/T=0.1\,$Hz/$\mu$K. Assuming a laser waist of $w=20\,\mu$m, and power $P=400\,$mW,
Fig.~\ref{Fig:5} illustrates the optical dipole potential, as well as its combination with a magnetic quadrupole field
$\textbf{B}=100\,\textrm{G/cm}\,\{-x/2,-y/2,z\}$ for atoms in the ground state $|F=2,m_F=2\rangle$, yielding an adiabatic magnetic potential
$U_B=\mu_B |\textbf{B}|$ \cite{general}.

In addition to the dramatically reduced heating rates afforded by blue-detuned (dark) dipole traps, it should be stressed \cite{KetterleSE} that dark
ring lattices should greatly reduce the observed heating due to micromotion of atoms during AOM beam scan in bright dipole traps. Additionally,
scattering a few photons tends to optically pump atoms into magnetically untrapped states: thus dark (bright) traps scatter more photons in the hottest
(coldest) parts of the potential leading to evaporative cooling (heating).

\section{Conclusions and Acknowledgments}
We have experimentally obtained reproducible optical ring lattices for use with blue-detuned light, which allow controllable rotation and spatial
modulation. The lattices produced are relatively insensitive to environmental conditions and do not require either LG beams, SLMs, or other complex
optics. This technique could possibly be utilised for rotation of a ``quantum register,'' and for new forms of optical tweezing.

%With AOM beam, even relatively incoherent light $(\Delta\nu>1\,$THz) can be used.

%hypocycloids, epicycloids, hypotrochoids, epitrochoids, archimedian spirals, lemniscates, and
%limaçons.

%"Standard technique" noise-eater.

%Taking advantage of quantum effects being most prevalent at low dimensionality, ring traps have been generated for quantum degenerate gases which are
%effectively equivalent to infinite 1D geometries [].

%Restress key advantages particularly precise control of arbitrary amplitude and phase modulation?
We gratefully acknowledge discussions with Dr. Malcolm Boshier and Dr. Ifan Hughes.

\end{document}